\documentclass[aps,prl,twocolumn,preprintnumbers,superscriptaddress]{revtex4}

\usepackage[dvips]{color}
\usepackage[normalem]{ulem}
\usepackage{amsmath}
\usepackage{enumerate}
\usepackage{amsfonts}
\usepackage{epsfig}

\newcommand{\be}{\begin{equation}}
\newcommand{\ee}{\end{equation}}
\newcommand{\bea}{\begin{eqnarray}}
\newcommand{\eea}{\end{eqnarray}}

 \newcommand{\bln}{\begin{align}}
\newcommand{\eln}{\end{align}}
\newcommand{\bst}{\begin{split}}
\newcommand{\est}{\end{split}}
\newcommand{\bi}{\begin{itemize}}
\newcommand{\ei}{\end{itemize}}
\newcommand{\bn}{\begin{enumerate}}
\newcommand{\en}{\end{enumerate}}

\def\lad{L} 
\def\lg{\lam_{GB}}
\def\ns{N_{\sharp}}
\def\tf{\tilde{f}}
\def\w{\tilde{\om}}
\def\q{\tilde{q}}

\def\ov{\over}
\def\le{\left}
\def\ri{\right}
\def\ha{{1\over 2}}
\def\lam{{\lambda}}

\def\al{{\alpha}}

\def \lam {\lambda}
\def \om {\omega}

\def\apr{{\alpha'}}
\newcommand{\p}{\partial}

\def\lam{{\lambda}}

\def\eeq{\end{equation}}

\begin{document}

\title {Viscosity Bound and Causality Violation}

\preprint{SU-ITP-08/02, MIT-CTP-3929}

\author{Mauro Brigante}
\affiliation{Center for Theoretical Physics,
 Massachusetts
Institute of Technology,
Cambridge, MA 02139
}

\author{Hong Liu}
\affiliation{Center for Theoretical Physics, Massachusetts
Institute of Technology,
Cambridge, MA 02139
}

\author{Robert C. Myers}
\affiliation{Perimeter Institute for Theoretical Physics, Waterloo, Ontario N2L 2Y5, Canada,\\
Department of Physics and Astronomy, University of Waterloo, Waterloo, Ontario
N2L 3G1, Canada
}

\author{Stephen Shenker}
\affiliation{Department of Physics, Stanford University, Stanford, CA 94305, USA
}

\author{Sho Yaida}
\affiliation{Department of Physics, Stanford University, Stanford, CA 94305, USA
}

\date{February, 2008}

\begin{abstract}
In recent work we showed that, for a class of conformal field theories (CFT) with
Gauss-Bonnet gravity dual, the shear viscosity to entropy density
ratio, $\eta/s$, could  violate the conjectured Kovtun-Starinets-Son viscosity bound,
$\eta/s\geq1/4\pi$. In this paper we argue, in the context of the
same model, that tuning $\eta/s$ below $(16/25)(1/4\pi)$ induces
microcausality violation in the CFT, rendering the theory inconsistent.
This is a concrete example in which inconsistency of a theory and a
lower bound on viscosity are correlated, supporting the idea of a
possible universal lower bound on $\eta/s$  for all consistent
theories.
\end{abstract}

\maketitle
\newpage


The anti-de Sitter (AdS)/conformal field theory (CFT)
correspondence~\cite{MaldacenaOriginal,GKP,Witten,WittenThermal}
has yielded striking insights into the dynamics of strongly
coupled gauge theories. Among them is the universality of the ratio of the
shear viscosity $\eta$ to the entropy density
$s$~\cite{Policastro:2001yc,Kovtun:2003wp,Buchel:2003tz,KSSbound}
\begin{equation}
\label{bound} \frac{\eta}{s} = \frac{1}{4\pi}
\end{equation}
for all gauge theories with an Einstein gravity dual in the limit $N
\to \infty$ and $\lam \to \infty$, where $N$ is the number of colors
and $\lam$ is the 't Hooft coupling. It was further conjectured
in~\cite{KSSbound} that~(\ref{bound}) is a universal lower bound
[the Kovtun-Starinets-Son (KSS) bound]  for all materials. So far, all known substances including water and liquid helium satisfy the bound. The systems
coming closest to the bound include the quark-gluon plasma created
at RHIC~\cite{Teaney:2003kp,rr1,songHeinz,Adare:2006nq,r2,Dusling:2007gi}
and certain cold atomic gases in the unitarity limit (see
e.g.~\cite{Schafer:2007ib}). $\eta/s$ for pure gluon QCD slightly
above the deconfinement temperature has also been calculated on the
lattice recently~\cite{Meyer:2007ic} and is about $30 \%$ larger
than~(\ref{bound}) (see also~\cite{sakai}). Furthermore, the leading
order $\apr$ correction to $\eta/s$ has been calculated for the dual
of type IIB string theory on $AdS_5 \times S^5$ and found to satisfy
the bound~\cite{BLS,Benincasa:2005qc}.
See~\cite{Cohen:2007qr,Cherman:2007fj,Chen:2007jq,Son:2007xw,Fouxon:2007pz}
for other discussions of the bound.

From the point of view of AdS/CFT, an interesting feature of the KSS
bound is that it is {\it saturated} by Einstein gravity. Thus at the
linearized order generic small corrections to Einstein gravity
violate the bound half of the time. Given that we do expect
corrections to Einstein gravity to occur in any quantum theory of
gravity, it appears that the bound is in immediate danger of being violated.
On the other hand, the correctness of the bound would
impose an important constraint on possible higher order corrections
to Einstein gravity.

Motivated by the vastness of the string landscape~\cite{landscape},
we have explored the modification of $\eta/s$ due to generic higher
derivative terms in the holographic gravity dual~\cite{us}. For
closely related work, including a plausible counterexample to the KSS bound, see~\cite{new}. In particular, for a class of
(3+1)-dimensional CFTs with Gauss-Bonnet gravity dual, described by
the classical action of the form~\cite{Zwiebach} (below $\Lambda =-
{6 \ov L^2}$ and the Gibbons-Hawking term~\cite{Myers} is suppressed)
\begin{eqnarray}
\label{action}
 I  &&= \frac{1}{16\pi G_N} \mathop\int d^{5}x \,
\sqrt{-g} \, [R-2\Lambda  \nonumber \\
 && +
{\lg \ov 2} \lad^2
(R^2-4R_{\mu\nu}R^{\mu\nu}+R_{\mu\nu\rho\sigma}R^{\mu\nu\rho\sigma})
] \ ,
\end{eqnarray}
we found that~\cite{us}
\begin{equation} \label{advertise}
\frac{\eta}{s}=\frac{1}{4\pi}[1- 4 \lg].
\end{equation}
We emphasize that this result is nonperturbative in $\lg$, not just
a linearly corrected value. From (\ref{advertise}) the KSS bound is
violated for $\lg > 0$ and as $\lg \rightarrow \frac{1}{4}$, the
shear viscosity goes to zero~\cite{footA}.

In this paper, we will argue that when $\lg>\frac{9}{100}$, the
theory violates  microcausality and is inconsistent. Thus, for (3+1)-dimensional CFT duals of
(4+1)-dimensional Gauss-Bonnet gravity, consistency of the theory
requires
 \be \label{newbound}
\frac{\eta}{s}\geq\frac{16}{25}\le(\frac{1}{4\pi}\ri).
 \ee
This provides a concrete example in which a lower bound on $\eta/s$
and the consistency of the theory are correlated. The 36\%
difference from the KSS bound is mysterious, and we discuss  two obvious
possibilities at the end. Our discussion below will rely
heavily on a few technical results derived in~\cite{us}, to which we
refer the readers for details and references.

The static black brane solution for (\ref{action}) can be written
as~\cite{Cai}
\begin{equation}
\label{bba} ds^2=-f(r)\ns^2dt^2
+\frac{1}{f(r)}dr^2+\frac{r^2}{\lad^2}
 \le(\mathop\sum_{i=1}^{3}dx_i^2 \ri),
\end{equation}
where
\begin{equation}
\label{perturb} f(r)=\frac{r^2}{\lad^2}\frac{1}{2\lg}
\le[1-\sqrt{1-4\lg\le(1-\frac{r_{+}^{\,4}}{r^4}\ri)} \ri] \ .
\end{equation}
In (\ref{bba}), $\ns$ is an arbitrary constant which specifies the
speed of light of the boundary theory. We will take it to be
 $ \ns^2 \equiv 
  \ha\le(1+\sqrt{1-4\lg}\ri)$
to set the boundary speed of light to unity. The horizon is located
at $r=r_{+}$ and the Hawking temperature is $T =\ns \frac{r_{+}}{\pi
\lad^2}$.
Such a solution describes the boundary theory on ${\bf R}^{3,1}$ at
a temperature $T$.

The shear viscosity can be computed by studying small metric
fluctuations $\phi = h^1_{\ 2}$ around the black brane
background~(\ref{bba}). We will take $\phi$ to be independent of
$x_1$ and $x_2$ and write
\begin{equation} \label{eorE}
\phi(t, \vec x, r)=
\mathop\int\frac{d\om dq}{(2\pi)^{2}}\,
\phi(r;\om,q)
\, e^{-i\omega t + i q x_3}  
\ .
\end{equation}
At quadratic level, the effective action for $\phi (r;\om,q)$ can be
found from (\ref{action}) as (up to surface terms)
 \be \label{efact}
 S = -\ha C \int dz  {d \om dq \ov (2 \pi)^2}
  \le(K |\p_z \phi|^2 - K_2 |\phi|^2  \ri)
 \ee
where $C$ is a constant, and
\bea \label{vds}
 K&=& z^2 \tf (z - \lg
 \tf'), \cr K_2 &=& K {\tilde \om^2 \ov \ns^2 \tf^2} - \tilde
 q^2 z \le(1- \lg \tf'' \ri) \ .
 \eea
Primes above denote derivatives with
respect to $z$ and we have introduced the following notation
 \bea \label{sDef}
&&z=\frac{r}{r_{+}},\ \ \tilde{\omega}=\frac{\lad^2}{r_{+}}\omega,\
\quad \tilde{q}=\frac{\lad^2}{r_{+}}q, \\
&& \tilde{f}=\frac{\lad^2}{r_{+}^2}f =  {z^2 \ov 2 \lg} \le(1 -
\sqrt{1-4 \lg + {4 \lg \ov z^4}} \ri).
 \eea
From (\ref{efact}) one can derive the retarded two-point function
for the stress tensor component $T_{12}$ in the boundary CFT and read off the shear
viscosity~(\ref{advertise}) in the small $q$ and $\om$ limit.
In~\cite{us} it was also observed that for $\lg > {9 \ov 100}$ and
sufficiently large $q$, the equation of motion for $\phi$ admits
solutions which can be interpreted as metastable quasiparticles of
the boundary CFT. We will now argue that these quasiparticles can
travel faster than the speed of light and thus violate causality.
We will first display this phenomenon using graviton geodesics.

\begin{figure}[t]
\vskip-0.25in
\includegraphics[scale=0.5]{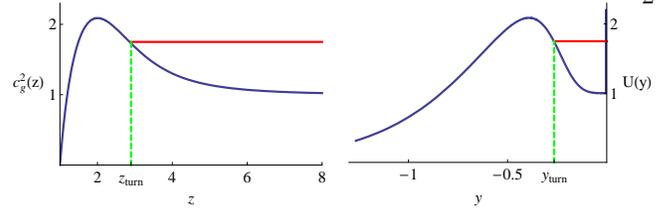}
\vskip-0.1in \caption{Left: $c_g^2 (z)$ as a function of $z$ for $\lg =
0.245$. $c_g^2$ has a maximum $c_{g,{\rm max}}^2$ at $z_{\rm max}$. As
$\lg$ is increased from $\lg={9 \ov 100}$ to $\lg = {1 \ov 4}$,
$c_{g,{\rm max}}^2$ increases from $1$ to $3$. $c_g^2 (z)$ also serves
as the classical potential for the 1D system (\ref{geoE}). The
horizontal line indicates the trajectory of a classical particle. Right:
 $U(y)$ [defined in (\ref{Upro})] as a function of $y$ for $\lg =0.245$.}
\label{fig:velo} \vskip-0.05in
\end{figure}

In a gravity theory with  higher derivative terms, graviton wave
packets in general do not propagate on the light cone of  a given
background geometry. The equation of motion following from~(\ref{efact}) can be written as
\be
\label{covariant}
{\tilde g}^{\mu\nu}_{\rm eff}{\tilde \nabla}_{\mu}{\tilde \nabla}_{\nu}\phi=0
\ee
where ${\tilde \nabla}_{\mu}$ is a covariant derivative with respect to the effective geometry ${\tilde g}_{\mu\nu}^{\rm eff}=\Omega^2 g_{\mu\nu}^{\rm eff}$ given by
 \be
\label{effgeoeq} g^{\rm eff}_{\mu\nu}dx^{\mu}dx^{\nu}
=f(r)\ns^2 \le(-dt^2 + {1 \ov c_g^2} dx_3^2 \ri)
+\frac{1}{f(r)}dr^2 \ .
 \ee
Here, $\Omega^2= \frac{K}{{\tilde f}}z(1-\lg{\tilde f}'')$ and
 \be \label{Nse}
c_g^2 (z)   = {\ns^2 \tilde f \ov z^2} {1-\lg \tf'' \ov 1 -
{\lg \tf' \ov z}}
 \ee
can be interpreted as the local ``speed of graviton'' on a constant
$r$-hypersurface. As pointed out in~\cite{us}, an important feature
of (\ref{Nse}) is that $c_g$ can be greater than $1$ when $\lg > {9
\ov 100}$ (see Fig.~\ref{fig:velo}). Note that $c_g$ has a
maximum $c_{g,{\rm max}}>1$ somewhere outside the horizon and approaches
$1$ near the boundary at infinity. We will restrict our attention to
$\lg > {9 \ov 100}$ for the rest of the paper.

From standard geometrical optics arguments~\cite{MTW}, in the large momentum limit, a localized
wave packet of a graviton should follow a null geodesic $x^{\mu}(s)$ in the
effective graviton geometry~(\ref{effgeoeq}). More explicitly, write the wave function
(\ref{eorE}) in the form $\phi=e^{i\Theta(t, r, x_3)}\phi_{en}(t, r, x_3)$
where $\Theta$ is a rapidly varying phase and $\phi_{en}$
denotes a slowly varying envelope function. Inserting into
(\ref{covariant}), we find at leading order
  \be \label{null}
  \frac{dx^{\mu}}{ds}\frac{dx^{\nu}}{ds}g_{\mu\nu}^{\rm eff}=0,
  \ee
with the identification $ {d x^\mu \ov ds}\equiv g_{\rm eff}^{\mu \nu} k_{\nu}\equiv g_{\rm eff}^{\mu \nu}\nabla_\nu \Theta$.
Given translational symmetries in the $t$ and $x_3$ directions, we can interpret $\om$ and $q$ as
conserved integrals of motion along the geodesic,
 \be  \label{cons}
 \om = \le(\frac{dt}{ds}\ri)f \ns^2 , \qquad
 q = \le(\frac{dx_3}{ds}\ri)f \ns^2 \frac{1}{c_g^2} \ .
 \ee
Assuming $q\ne0$ and rescaling the affine parameter as $\tilde{s}=q
s/\ns$, we get from (\ref{null}) and (\ref{cons})
 \be \label{geoE}
\le(\frac{dr}{d\tilde{s}}\ri)^2=\al^2-c_g^2, \ \ \ \al
\equiv\frac{\om}{q}.
 \ee
This describes a one-dimensional particle of energy $\al^2$ moving in
a potential given by $c_g^2$. As is clear from Fig.~\ref{fig:velo},
geodesics starting from the boundary can bounce back to the
boundary, with a turning point $r_{\rm turn} (\al)$ given by
 \be \label{turn}
 \alpha^2 = c_g^2 (r_{\rm turn}) \ .
 \ee
In contrast,  for $\lg\leq\frac{9}{100}$, $c_{g} (z)$ is a
monotonically increasing function of $z$ and there is no bouncing
geodesic. For a null bouncing geodesic starting and ending at the
boundary, we then have
 \be \label{timeelapsed}
\Delta t(\alpha)=2\int_{r_{\rm turn}(\alpha)}^{\infty}
\frac{\dot{t}}{\dot{r}}dr=\frac{2}{\ns}
\int_{r_{\rm turn}(\alpha)}^{\infty}\frac{\alpha}{f
\sqrt{\alpha^2-c_g^2}}dr,
 \ee
 \be \label{distanceelapsed}
\Delta x_3(\alpha)=2\int_{r_{\rm turn}(\alpha)}^{\infty}
\frac{\dot{x_3}}{\dot{r}}dr=\frac{2}{\ns}
\int_{r_{\rm turn}(\alpha)}^{\infty}\frac{c_g^2}{f
\sqrt{\alpha^2-c_g^2}}dr,
 \ee
where dots indicate derivatives with respect to $\tilde{s}$.

In the boundary CFT we have local operators which create bulk disturbances at 
infinity that propagate on graviton geodesics sufficiently deep inside the bulk ($r\lesssim\omega$)~\cite{Polchinski}.
In particular, we expect microcausality violation in the boundary CFT if there exists a
bouncing graviton geodesic with $\frac{\Delta x_3(\alpha)}{\Delta
t(\alpha)}>1$~\cite{footC}. Now, as $r_{\rm turn} \rightarrow
r_{\rm max}$ ($\alpha \rightarrow c_{g,{\rm max}}$), a geodesic hovers near
$r_{\rm max}$ for a long time, propagating with a speed $c_{g,{\rm max}}$ in
$x_3$-direction. Indeed, the integrals in (\ref{timeelapsed}) and
(\ref{distanceelapsed}) are dominated by contributions near
$r_{\rm max}$. In such a limit, the ratio of the integrand in $\Delta
x_3(\alpha)$ to that in $\Delta t(\alpha)$ near $r_{\rm max}$ is $c_{g,{\rm max}}$. Thus, $\frac{\Delta x_3(\alpha)}{\Delta t(\alpha)}\rightarrow
c_{g,{\rm max}}>1$, violating causality.

We will now show explicitly that the superluminal graviton propagation
described above corresponds to superluminal propagation of metastable
quasiparticles~\cite{footD} in the boundary CFT with ${\Delta x_3 \ov \Delta t}$ identified as
the group velocity of the quasiparticles. For this purpose, we rewrite the full wave equation (\ref{covariant}) in a Schr\"{o}dinger form
 \be \label{enr}
 - \p_y^2 \psi + V(y) \psi = \w^2 \psi
 \ee
with $\psi$ and $y$ defined by
 \be \label{hockeystick}
{dy \ov dz} = {1 \ov \ns \tf(z)} , \qquad \psi = B \phi , \qquad B
= \sqrt{K \ov \tf} \ ,
 \ee
and
\be \label{potential}
 V (y) = \q^2 c_g^2 (z) + V_1,  \quad  V_1 (y) 
= {\ns^2 \tf^2 \ov B} \le(B'' + {\tf' \ov \tf} B' \ri) \ .
 \ee
In the above primes denote derivatives with respect to $z$.
Note that $y(z)$ is a monotonically increasing function of $z$ with $y
\rightarrow 0$ as $z \rightarrow \infty$ (boundary) and $y
\rightarrow -\infty$ as $z \rightarrow 1$ (horizon). $c_g^2(z)$ is
given by (\ref{Nse}). $V_1$ is a monotonically increasing function
of $y$ (for $\lg>0$) with $V_1(y=-\infty)=0$ and $V_1\sim y^{-2}$ as
$y \rightarrow 0$.

\begin{figure}[t]
\vskip-0.25in
\includegraphics[scale=0.4]{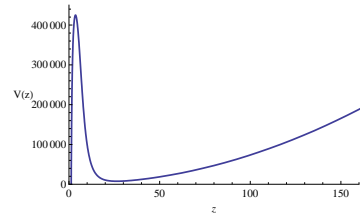}
\vskip-0.1in
\caption{$V(z)$ as a function of $z$ for $\lg =0.2499$ and $\q =500$.
} \label{fig:P}
\vskip-0.05in
\end{figure}

Since $c_g^2$ is monotonically decreasing for $r > r_{\rm max}$, for
large enough $\q$, $V(y)$ develops a well and admits metastable
states (see Fig.~\ref{fig:P}). The wave functions of such metastable
states are normalizable at the AdS boundary and have an in-falling tail at the horizon, corresponding to quasiparticles in the boundary CFT~\cite{footD}.


Now consider the limit $\q \to \infty$. Since $V_1$ is independent of
$\q$, the dominant contribution to the potential is given by $\q^2 c_g^2 (z)$ except for a tiny region
$y \gtrsim -{1 \ov \q}$. Thus in this limit, we can simply replace
$V_1 (y)$ by $V_1 (y) = 0$ for all $y < 0$ and $V_1 (0) = +\infty$.
Equation
(\ref{enr}) can then be written as
 \be  \label{sinE}
 -\hbar^2 \p_y^2 \psi + U (y) \psi = \al^2 \psi, \qquad \hbar \equiv {1 \ov \q} \to 0
 \ee
 where $\al$ was introduced in~(\ref{geoE}) and (see Fig.~\ref{fig:velo})
  \be \label{Upro}
  U(y) = \begin{cases} c_g^2 (y) & y < 0 \cr
                         + \infty & y=0
                         \end{cases}   \  .
                         \ee
In the $\hbar \to 0$ limit, we can apply the WKB approximation.
The leading WKB wave function $e^{i\Theta(t, r, x_3)}$ is just the rapidly varying phase of the geometric optics approximation.
The real part of $\al^2$ satisfies the Bohr-Sommerfeld quantization condition (with $n$ some integer)
 \be \label{bors}
 \tilde q \int_{y_{\rm turn}}^0 dy \, \sqrt{\al^2 - c_g^2 (y)} = (n-{1 \ov 4} ) \pi
 \ee
The above equation determines $\om$ as a function of $q$ for each given $n$. Taking the derivative with respect to $q$ on both sides of (\ref{bors}), we find that the group velocity of the
quasiparticles is given by
 \be
 v_g = {d \om \ov d q} = \frac{\Delta x_3 (\alpha)}{\Delta t (\alpha)}
 \ee
where $\Delta t (\alpha)$ and $\Delta x_3 (\alpha)$ are given by (\ref{timeelapsed}) and (\ref{distanceelapsed}) respectively. Thus as argued in the paragraph below (\ref{distanceelapsed}), $v_g$ approaches $c_{g,{\rm max}}> 1 $ as $\al \to c_{g,{\rm max}}$, violating causality.
In this limit the WKB wave function is strongly peaked near $r_{\rm max}$, reflecting the long time the geodesic spends there.
One can also estimate the imaginary part of $\al^2$ (or $\om$), which has the form $e^{-h(\al) \q}$ with $h(\al)$ given by the standard WKB formula. Thus in the $\q \to \infty$ limit the quasiparticles become stable. Presumably local boundary operators that couple primarily to the long-lived quasiparticles can be constructed by following~\cite{Polchinski}.

To summarize, we have argued that signals in the boundary theory
propagate outside the light cone. In a boosted frame disturbances
will propagate backward in time. Since the boundary theory is
nongravitational, these are unambiguous signals of causality
violation and hence inconsistency.

Here we observe causality violation in the high
momentum limit. This is in agreement with the
expectation that causality should be tied to the local,
short-distance behavior of the theory. Also, a sharp transition from
causal to acausal behavior as a function of $\lg$ is possible
because of the limiting procedure $\q\rightarrow\infty$ needed in
our argument. A more rigorous derivation of these phenomena using the full spectral function obtained from the Schr\"{o}dinger operator would be desirable.

We argued that, for a (4+1)-dimensional Gauss-Bonnet gravity,
causality requires $\lg\leq\frac{9}{100}$. Thus, consistency of this
theory requires,
 \be
\frac{\eta}{s}\geq \frac{16}{25}\le(\frac{1}{4\pi}\ri).
 \ee
This still leaves rooms for a violation of the KSS bound. We see two
possibilities.

First, it could be that Gauss-Bonnet theory with
$\lg\leq\frac{9}{100}$ is consistent and appears as a classical
limit of a consistent theory of quantum gravity, somewhere in the
string landscape (see~\cite{new} for a plausible counterexample to the KSS bound). Maybe this is how nature works and the KSS
bound can be violated, at least by 36\%.

Alternatively, it could be that there is a more subtle inconsistency in the theory within the window of $0<\lg\leq\frac{9}{100}$. These issues
deserve further investigation.

We thank D.~T.~Son, A.~Starinets, and L.~Susskind for discussions.
MB and HL are partly supported by the U.S. Department of Energy
(D.O.E) under cooperative research agreement \#DE-FG02-05ER41360. HL
is also supported in part by the A.P. Sloan Foundation and the U.S.
Department of Energy (DOE) OJI program. Research at Perimeter
Institute is supported by the Government of Canada through Industry
Canada and by the Province of Ontario through the Ministry of
Research \& Innovation. RCM also acknowledges support from an NSERC
Discovery grant and from the Canadian Institute for Advanced
Research. SS is supported by NSF grant 9870115 and the Stanford
Institute for Theoretical Physics. SY is supported by an Albion
Walter Hewlett Stanford Graduate Fellowship and  the Stanford
Institute for Theoretical Physics.


\begin{thebibliography}{11}
\bibitem{footA}
  As discussed in~\cite{us},
$\lg$ is bounded above by $\frac{1}{4}$ for the theory to have a
boundary CFT, and $\eta/s$ never decreases below $0$.
\bibitem{footC}
To be precise this only indicates the presence of a pole outside the 
boundary CFT light cone in the time-ordered two-point function.
To be complete, we need to show that the retarded two-point 
function does not vanish outside the light cone.
\end{thebibliography}
\end{document}